# CloudBrain-ReconAI: A Cloud Computing Platform for MRI Reconstruction and Radiologists' Image Quality Evaluation

Yirong Zhou, Chen Qian, Jiayu Li, Zi Wang, Yu Hu, Biao Qu, Liuhong Zhu, Jianjun Zhou, Taishan Kang, Jianzhong Lin, Qing Hong, Jiyang Dong, Di Guo, and Xiaobo Qu

*Abstract*—Efficient collaboration between engineers and radiologists is important for image reconstruction algorithm development and image quality evaluation in magnetic resonance imaging (MRI). Here, we develop CloudBrain-ReconAI, an online cloud computing platform, for algorithm deployment, fast and blind reader study. This platform supports online image reconstruction using state-of-the-art artificial intelligence and compressed sensing algorithms with applications for fast imaging (Cartesian and non-Cartesian sampling) and high-resolution diffusion imaging. Through visiting the website, radiologists can easily score and mark images. Then, automatic statistical analysis will be provided. CloudBrain-ReconAI is now open accessed at https://csrc.xmu.edu.cn/CloudBrain.html and will be continually improved to serve the MRI research community.

## I. INTRODUCTION

MAGNETIC Resonance Imaging (MRI) reconstruction algorithms have been developed fast to accelerate imaging speed or enhance image quality. Typical algorithms include optimization [1-6] and deep learning approaches [7-14]. However, image quality is still a big concern for clinical applications [15]. Although some objective criteria, such as peak-signal-to-noise-ratio [16], mean square error, and image structure loss [17], are widely employed. Subjective evaluations by experienced radiologists are also very valuable. In addition, radiologists may be busy with clinical tasks thus an efficient evaluation system is highly expected.

To the best of our knowledge, there are few integrated platforms for easy MRI reconstruction and radiologists' evaluation. Michael Lustig's group has developed the Berkeley Advanced Reconstruction Toolbox (BART) [18], which is a free and open-source image-reconstruction framework. Tobias Knopp's group [19] has proposed MRIReco.jl, which is written under the Julia framework and achieves reconstruction speeds that are comparable to popular C/C++ libraries. Both platforms have excellent performance and bring convenience to MRI researchers. Still, some limitations exist, such as less support for deep learning algorithms, radiologists' evaluations, and relatively complex installation and environment configurations.

In this work, we developed the CloudBrain-ReconAI platform (Fig. 1), a cloud computing-based MRI reconstruction and image quality evaluation system, to provide efficient collaborative services for experts in engineering, clinics, and other fields. This platform aims to accelerate MRI reconstruction algorithm development and facilitate clinical applications.

CloudBrain-ReconAI supports the deployment of multiple optimizations and deep learning algorithms (Fig. 2). For fast MRI under Cartesian sampling [20,21], compressed sensing-based pFISTA-SENSE [4] and model-based deep learning methods, e.g. pFISTA-NET [12], DOTA [22], and VarNET [23], are deployed on our platform. For fast MRI under 2D non-Cartesian sampling, e.g. pseudo golden angle radial sampling [24], the deep learning reconstruction method pFISTA-AI [25], which outperforms the state-of-the-art NC-PDNet [26], and another compressed sensing method pFISTA [27], are compared online. For fast MRI under 3D non-Cartesian sampling, e.g. the golden-angle sampling for dynamic contrast-enhanced (DCE) imaging [28, 29], deep learning reconstruction method stDLNN [30], and one compressed sensing reconstruction method pFISTA [27] are also compared on the cloud. For high-resolution diffusion MRI (DWI), POCS-ICE (PICE) [31], a low-rankness-based optimization PAIR [32], and a deep learning method PIDD [33, 34] are also successfully

This work was supported by the National Natural Science Foundation of China (62122064, 62331021 and 62371410), the Natural Science Foundation of Fujian Province of China (2023J02005), Industry-University Cooperation Projects of the Ministry of Education of China (231107173160805), National Key Research and Development Program of China (2023YFF0714200), the President Fund of Xiamen University (20720220063), and the Nanqiang Outstanding Talents Program of Xiamen University. (Yirong Zhou and Chen Qian Contributed equally to this work.) (*Corresponding author: Xiaobo Qu, Email: quxiaobo@xmu.edu.cn)

Yirong Zhou, Chen Qian, Jiayu Li, Zi Wang, Hu Yu, Jiyang Dong, and Xiaobo Qu are with the Department of Electronic Science, Biomedical Intelligent Cloud R&D Center, Fujian Provincial Key Laboratory of Plasma and Magnetic Resonance, National Institute for Data Science in Health and Medicine, Xiamen University, Xiamen 361102, China.

Biao Qu is with the Department of Instrumental and Electrical Engineering, Xiamen University, Xiamen 361005, China.

Liuhong Zhu and Jianjun Zhou are with the Department of Radiology, Zhongshan Hospital (Xiamen), Fudan University, Xiamen 361015, China.

Taishan Kang and Jianzhong Lin are with the Department of Radiology, Zhongshan Hospital Xiamen University, School of Medicine, Xiamen University, Xiamen 361004, China.

Hong Qing is with the China Mobile Group, Xiamen 361000, China.

Di Guo is with the School of Computer and Information Engineering, Xiamen University of Technology, Xiamen 361024, China.

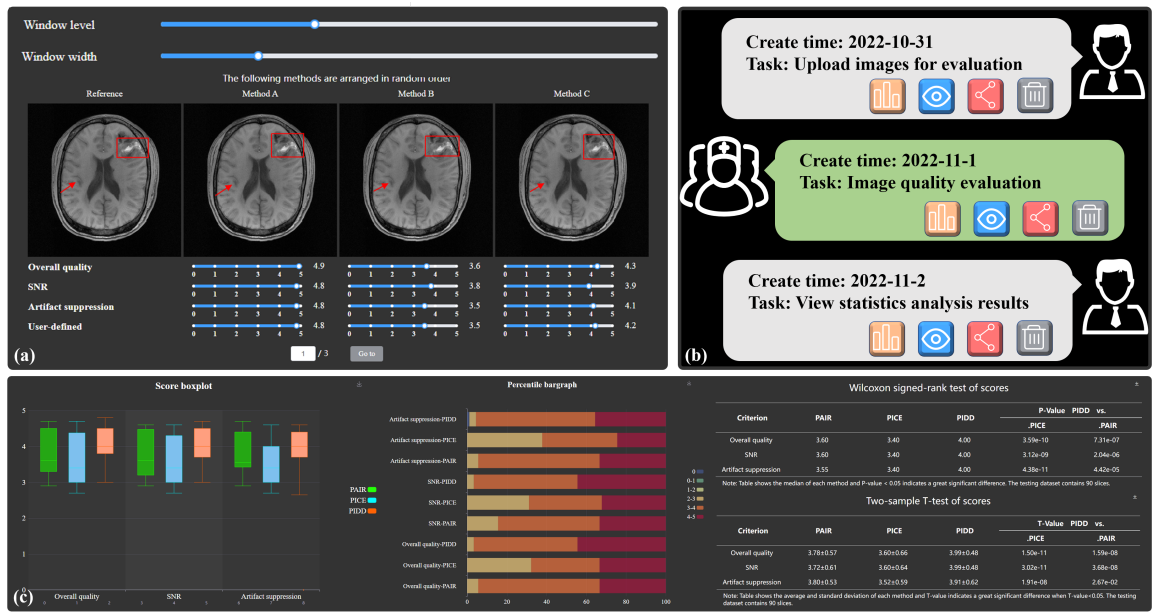

**Fig. 1.** Whole interface of CloudBrain-ReconAI. (a) Online scoring, (b) collaborative working, and (c) statistical analysis.

deployed.

For the subjective image quality assessment, a multi-reader online collaboration mode is provided. To facilitate evaluation, many supplementary tools, such as labelling a region of interest, adjusting the window width and window level for better clinical inspection, and automatically generating charts of statistical analysis for evaluation scores. To better serve the MRI research community, this platform is free and expected to incorporate more radiologists' knowledge into advanced MRI reconstruction algorithm design.

In the rest of this paper, the service, architecture, and application examples will be provided.

## II. SERVICES ON CLOUDBRAIN-RECONAI

CloudBrain-ReconAI provides a clear workflow for the MRI reconstruction and image quality evaluation (Fig. 2).

*A. Whole Workflow*

The platform is freely accessed at https://csrc.xmu.edu.cn/ReconAndEval.html. Users can register an account via the email address or use our test accounts (account1: researcher1, password: researcher1! and account2: radiologist1, password: radiologist1! and account3: radiologist2, password: radiologist2!) The user manual and demo data are also provided on the homepage for a quick try.

After registration, the whole workflow includes five steps:

1) Pre-process k-space data into the required ISMRM Raw Data format [35], a vendor neutral standard for describing data from magnetic resonance acquisitions and reconstructions, (refer to the user manual for details) and upload them to the platform. Not limited to ISMRM Raw Data format, we also support other k-space data formats, including MATLAB ".mat" format and other raw data files from Philips and Siemens.

2) Select reconstruction algorithms and appropriate parameters to reconstruct the uploaded data.

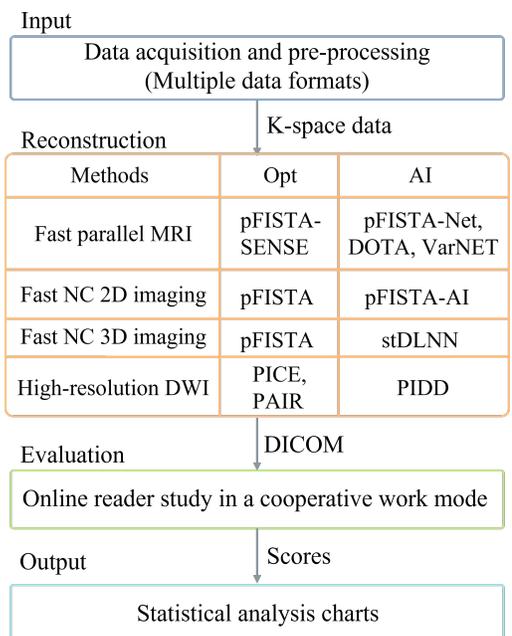

**Fig. 2.** Workflow on CloudBrain-ReconAI. Note: Opt, AI, and NC are short for optimization, artificial intelligence, and non-Cartesian sampling, respectively.

3) Check the reconstructed images and send them to accounts of pre-registered radiologists for image quality evaluation (Steps 1 and 2 can be skipped if the reconstructed images in DICOM format have been uploaded to the server).

4) Radiologists independently perform reader studies on reconstructed images according to the pre-set evaluation criteria by scoring and labelling. This platform also supports other imaging modalities, such as CT, X-ray, and ultrasound, saved in DICOM format, enabling evaluation with more information.

5) Evaluation scores from all radiologists will be analyzed and four statistical analysis charts will be generated automatically.

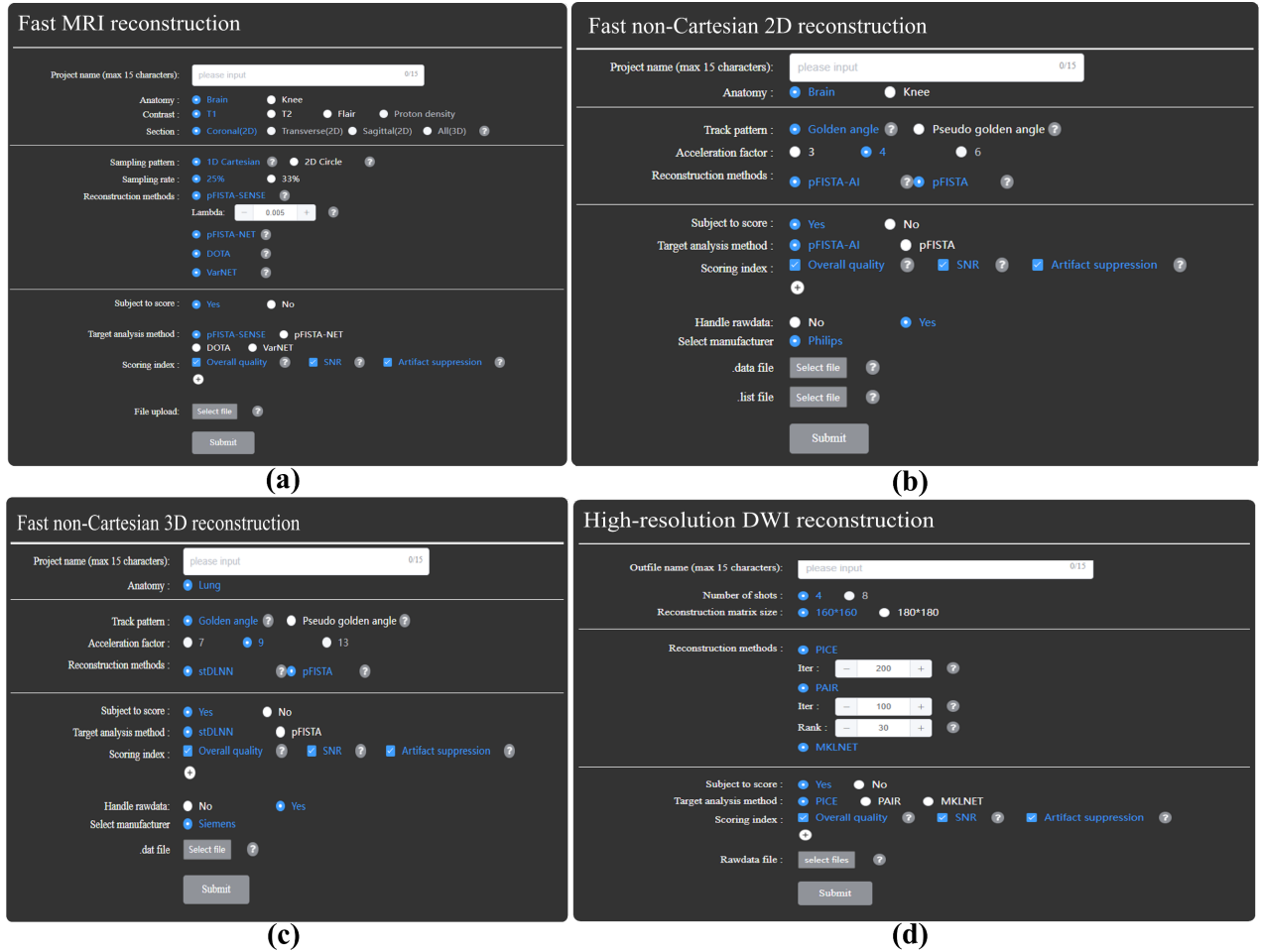

**Fig. 3.** Four image reconstruction tasks. (a) Screenshot of parameters selection page for fast MRI reconstruction, (b) screenshot of parameters selection page for fast non-Cartesian 2D reconstruction, (c) screenshot of parameters selection page for fast non-cartesian 3D reconstruction, and (d) screenshot of parameters selection page for high-resolution DWI reconstruction.

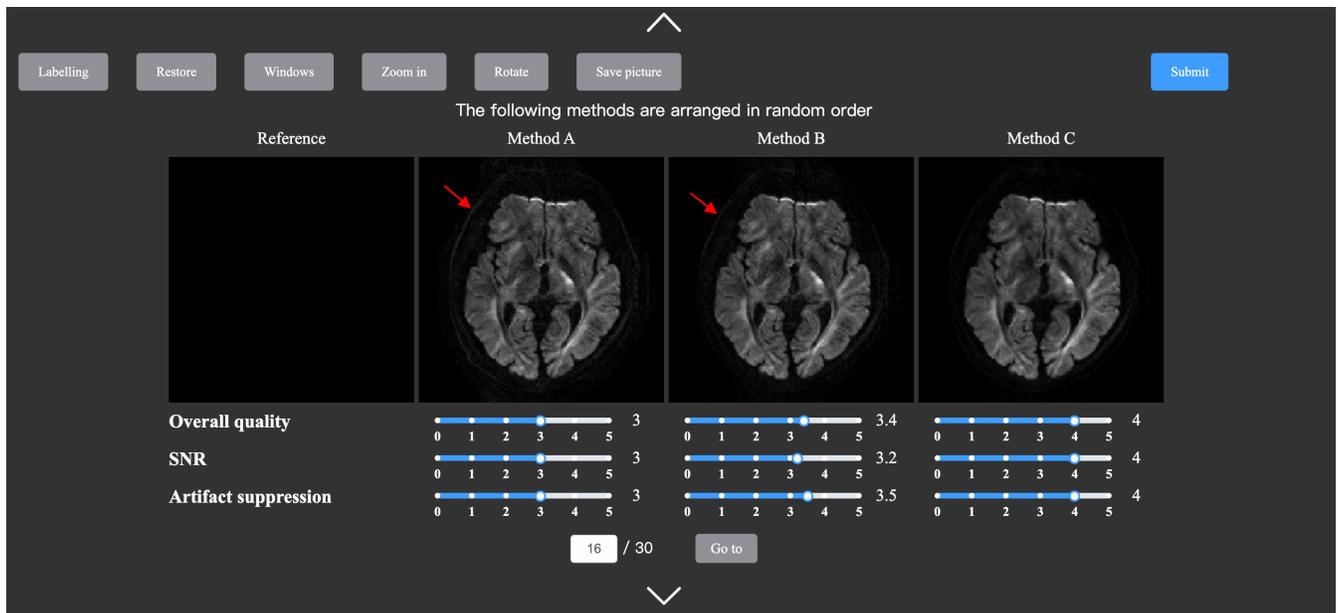

**Fig. 4.** The screenshot of the image quality evaluation system. Overall quality, artifacts, signal to noise ratio are the three default metrics for evaluation. Six functional icons including labelling, restore, windows, zoom in, rotate, and save pictures are shown in the top left corner.

*B. MRI Reconstructions*

Two MRI reconstruction applications, including the fast MRI and high-resolution diffusion MRI, are provided.

For fast MRI reconstruction under Cartesian undersampling, image artifacts must be removed. Both pFISTA-SENSE [4], pFISTA-Net [12], DOTA [22], and VarNET [23] have been deployed (Fig. 3(a)). The pFISTA-SENSE is a compressed sensing approach that iteratively reconstructs an image by enforcing its sparsity under the tight frame redundant representation [4, 36]. This approach was evidenced to be fast and remove the image artifacts well. pFISTA-Net is an unrolled deep learning network inspired by the former [4]. In pFISTA-Net, the fixed sparsity transform, e.g. wavelets, is replaced by learnable convolutional filters, making the algorithm a better fit for a specific MRI image database. DOTA first applies a 1D inverse Fourier transform on the frequency encoding dimension and subsequently transforms it into a reconstructed image using deep neural networks. VarNet embeds the concept of generalized compressed sensing into a deep learning framework using an unrolled gradient descent method and formulates it as a variational model.

For fast MRI reconstruction under non-Cartesian undersampling (2D radial sampling in Fig. 3(b) and 3D radial sampling in Fig. 3(c)), the methods pFISTA [27], pFISTA-AI [25], and stDLNN [30] have been deployed. pFISTA supports sparse image reconstruction for both static and dynamic imaging. It provides a theoretical convergence condition of the projected fast iterative soft-thresholding algorithm (pFISTA) for the tight frame sparse reconstruction. pFISTA-AI incorporates a deep unrolled neural network of pFISTA for static radially sampled MRI image reconstruction. The stDLNN is an innovative framework that combines spatial-temporal dictionary learning with deep neural networks to accelerate dynamic MRI.

For high-resolution DWI reconstruction, PICE [31], PAIR [32], and PIDD [33, 34] have been deployed (Fig. 3(d)). PICE and PAIR are two optimization-based methods. PICE alternatively updates the shot phase and magnitude images in an iterative process. PAIR simultaneously regularizes the reconstruction of the shot phase and image magnitude. PIDD is a physics-informed deep diffusion MRI reconstruction method [37], including the data synthesis of multi-shot DWI and the reconstruction with a deep learning network.

Without having to run any programming code on the user side, a friendly user interface is provided on the website thus one can easily use these algorithms. Essential algorithm parameters are listed on the parameter selection page. Users can quickly learn their effects by clicking the question icon next to the parameters (Fig. 3).

*C. Image Quality Evaluation*

Image quality assessment is another key step on the platform.

The system supports a multi-user collaboration mode. After loading the online reconstructed image or uploading offline reconstructed images, algorithm developers invite pre-registered radiologists to perform a reader study. Then, radiologists score and analyze the images independently.

CloudBrain-ReconAI provides three clinically concerned metrics of images as default, including overall quality, artifact suppression, and signal-to-noise ratio (SNR). Extra customized metrics such as the contrast-to-noise ratio and the lesion visibility can be added (Fig. 5). All the criteria follow the Likert scale (See more details in Table I) [38-41].

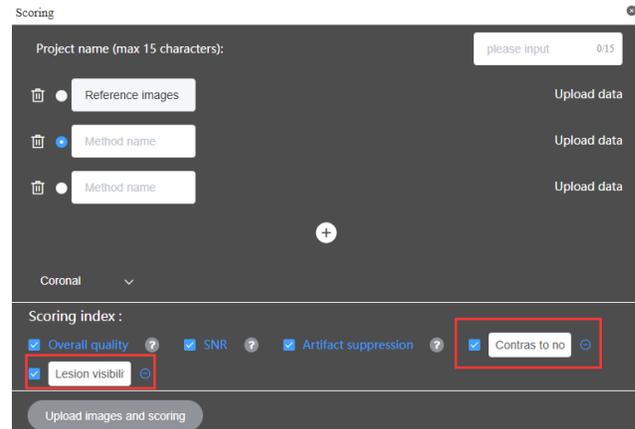

**Fig. 5.** Added scoring of contrast-to-noise ratio and lesion visibility.

To make evaluations easy and convenient, many small tools have been designed (Fig. 4). The labelling icon contains a set of several annotation tools, e.g. arrows and selection boxes, to mark the regions of interest such as lesions. These marks will be automatically saved and can be redisplayed by clicking the restore icon. The windows icon includes window width and window level to better demonstrate certain anatomy or pathology. The window width is the range of the grayscale that can be displayed. The center of the grayscale range is referred to as the window level. The user can simultaneously adjust the window width and window level of multiple images using two adjustable sliders or adjust these parameters for a single image by mouse-clicking on the target image. The window level adjusting is conducted by rolling the mouse up and down while the window width adjusting is performed by rolling the mouse left and right. The zoom-in icon enables users to enlarge the pictures to study image details. The rotate button helps to rotate a picture in the wrong direction back to the correct direction. To avoid evaluation the bias of reader evaluation, for each slice, the method name is removed and the order of reconstructed images obtained by different methods is randomized.

According to the selected evaluation metrics, radiologists need to give each image (or frame) a score. The score of every metric for each image (or frame) is provided by dragging or clicking on a slider which has a range from 0 to 5 with a precision of 0.1 (0~1: Poor, 1~2: Low, 2~3: Medium, 3~4: Good, 4~5: Excellent). To avoid bias in the evaluation of methods, radiologists are blind to these methods and the order of methods on each slice is randomly placed.

After finishing the evaluation and submitting the results, scores from all assigned accounts will be collected for further statistical analysis. All raw scores and statistical charts can be

TABLE I
5-POINT LIKERT SCALES SCORING CRITERIA

| Metric | Poor (Score=0~1) | Low (Score=1~2) | Medium (Score=2~3) | Good (Score=3~4) | Excellent (Score=4~5) |
|---|---|---|---|---|---|
| Overall quality | Significant issues | Noticeable issues | Minor issues | Minimal issues | No issues |
| Signal-to-noise ratio | Very weak signal | Weak signal, noisy | Visible signal, some noise | Strong signal, minimal noise | Very strong signal, no noise |
| Artifact suppression | Serious artifacts | Noticeable artifacts | Minimized artifacts | Well suppressed | Effectively suppressed |
| Contrast-to-noise ratio | Very low contrast, high noise | Low contrast, noisy | Medium contrast | High contrast | Very high contrast |
| Lesion visibility | Very difficult | Obscured by noise | Minor obstructions | Clear with minimal issues | Extremely clear |

downloaded for post-processing. Currently, the percentage plot, box plot, two-sample T-test [42], and Wilcoxon signed-rank test [43] have been adopted. The percentage and box plots analyze the score distribution, such as the median, mean, and standard deviation. The two-sample T-test and Wilcoxon signed-rank test evaluate the significant differences between reconstruction methods. Their mathematical formulations are summarized below:

**Two-sample T-test** determines if two population means are equal [42]. To explain the calculation, we take the overall quality metric shown in Fig. 6 as an example.

| Criterion | stDLNN | pFISTA | T-Value stDLNN vs. pFISTA |
|---|---|---|---|
| Overall quality | 4.53±0.16 | 4.12±0.43 | 5.65e-26 |
| SNR | 4.53±0.18 | 4.06±0.40 | 4.74e-30 |
| Artifact suppression | 4.53±0.18 | 4.17±0.41 | 3.68e-23 |
| Contrast to noise | 4.52±0.18 | 4.05±0.48 | 1.56e-26 |
| Lesion visibility | 4.51±0.22 | 4.18±0.33 | 1.91e-27 |

**Fig. 6.** Two-Sample T-test statistic results of overall quality.

To calculate statistic $T$, each subject has $N$ slices, which are evaluated by $M$ radiologists.
**Step 1** Calculate the difference:
$$d_i = x_i - y_i, \ i = 1, 2, 3, \ldots, N \times M, \tag{1}$$
where $x_i$ and $y_i$ are the evaluation scores from two methods, and $d_i$ is the difference between $x_i$ and $y_i$.

**Step 2** Calculate the mean $\bar{d}$ and standard deviation $s_d$ of $d_i$:
$$\bar{d} = \frac{1}{N \times M} \sum_{i=1}^{n} d_i, \ s_d = \sqrt{\frac{1}{N \times M - 1} \sum_{i=1}^{N \times M} (d_i - \bar{d})^2}. \tag{2}$$

**Step 3** Calculate the statistic $T$:
$$T = \frac{\bar{d}}{s_d / \sqrt{N \times M}}. \tag{3}$$

**Step 4** Get the $P$ value:
$$P(T) = \frac{Gama(\frac{v+1}{2})}{\sqrt{vp}\,Gama(\frac{v}{2})}(1+\frac{T^2}{v}), \ Gama(v) = \int_0^\infty T^{v-1} e^{-T} dT, \tag{4}$$

where $v = (N \times M - 1)$ is the freedom of statistic $T$, $Gama(v)$ is the gamma function, and $P(T)$ is the $P$ value.

Taking the statistical criteria of overall quality as an example, $N$= 66, $M$= 3, $\bar{d}$ = 0.41, $s_d$ = 0.47, $T$= 12.24, $v$ = 197. We can calculate the $P$ value is 5.56 × 10$^{-26}$, which is significantly less than 0.05, indicating that the method stDLNN has great significance with the method pFISTA.

**Wilcoxon signed-rank test** is a non-parametric test used to assess whether two paired samples come from populations with the same distribution [43]. We use the overall quality metric shown in Fig. 7 as an example.

| Criterion | stDLNN | pFISTA | P-Value stDLNN vs. pFISTA |
|---|---|---|---|
| Overall quality | 4.50 | 4.10 | 1.19e-25 |
| SNR | 4.50 | 4.00 | 3.60e-25 |
| Artifact suppression | 4.50 | 4.20 | 1.52e-24 |
| Contrast to noise | 4.50 | 4.00 | 2.46e-25 |
| Lesion visibility | 4.50 | 4.20 | 1.80e-22 |

**Fig. 7.** Wilcoxon Signed-Rank Test statistic results of overall quality.

To calculate statistic $W$, each subject has $N$ slices, which are evaluated by $M$ radiologists.
**Step 1** Calculate the difference:
$$d_i = x_i - y_i, \ i = 1, 2, 3, \ldots, N \times M, \tag{5}$$
where $x_i$ and $y_i$ are the evaluation scores from two methods, and $d_i$ is the difference between $x_i$ and $y_i$.

**Step 2** Calculate the $d_j^*, j = 1, 2, 3, \ldots, J(J \le N \times M)$ which is removed the zero value of $d_i$, and $J$ is the slice number after removing the zero.

**Step 3** Compute the absolute value $|d_j^*|$ of $d_j^*$.

**Step 4** Sort $|d_j^*|$, and use this sorted list to define ranks $R(|d_j^*|)$. The rank of the smallest observation is one, the rank of the next smallest is two, and so on.

**Step 5** Calculate the sum of positive rank $W^+$ and negative rank $W^-$:
$$W^+ = \sum_{j, d_j^* > 0} R(|d_j^*|), \ W^- = \sum_{j, d_j^* < 0} R(|d_j^*|). \tag{6}$$

**Step 6** Calculate the statistic $W$:
$$W = \min(W^+, W^-). \tag{7}$$

**Step 7** Calculate the $Z$ score:

$$Z = \frac{W - \frac{J(J+1)}{4}}{\sqrt{\frac{J(J+1)(2J+1)}{24}}}. \quad (8)$$

**Step 9** Get the *P* value:

$$P = 2 \times \Phi(-|Z|), \ \Phi(Z) = \frac{1}{\sqrt{2p}} \int_{-\infty}^{Z} e^{-\frac{t^2}{2}} dt, \quad (9)$$

where $\Phi(Z)$ denotes the cumulative distribution function.

Taking the statistical criteria of overall quality as an example, $N = 66$, $M = 3$, $W = 528.50$, $J = 198$, $Z = -11.54$. We can calculate the *P* value is $1.19 \times 10^{-25}$, which is significantly less than 0.05, indicating that the method stDLNN has great significance with the method pFISTA.

## III. ARCHITECTURES OF CLOUDBRAIN-RECONAI

This section introduces the workflow and system architecture, including the hardware resource, security, and privacy.

*A. System Architecture*

CloudBrain-ReconAI adopts the browser/service architecture (Fig. 8), which mainly consists of three parts: browser, service, and data access layer (DAL).

**Browser:** With a common browser, e.g. Chrome, the website can be easily accessed to obtain services without downloading or installing additional software. The website has a user-friendly interface with many functional buttons. Each button is a visualization of a certain application programming interface (API). By clicking the corresponding button, the service will be automatically called.

**Service:** Service requests from the browser will be transmitted to servers through the Nginx gateway. Three servers are employed in master-slave mode. The master provides reconstruction and image quality evaluation services. The slavers are equipped with reconstruction services to relieve the pressure on the master when the workload is heavy. The servers employ Google remote procedure calling for communication and network file system for data sharing.

**DAL:** All data are stored on the DAL with an effective strategy, such as MySQL, Redis, and MongoDB. MySQL stores structural data, such as the user name and password. Redis saves frequently used data such as authentication information. Other non-relational and big data, e.g. unreconstructed k-space data and reconstructed images, are stored in MongoDB.

*B. Hardware Resource and Performance*

To increase the usability and user experience, CloudBrain-ReconAI takes comprehensive measures to manage the hardware resource. The performance of data uploading, image reconstructions, and image loading is introduced below.

The uploading and downloading speed greatly affect the user experience. MRI k-space raw data with high dimensionality usually has a large size, reaching the gigabyte level. Splitting large files into multiple smaller files (default size is 4 MB) for parallel transfer is employed to improve uploading speed. Then, transfer requests are sent to a cluster of servers in parallel. When all the file slices are received, the server will merge these files into the original file, and verify the MD5 of the merged file and origin file to ensure integrity. We currently support two versions: public and private networks. For the public network on the internet, the uploading time of 1GB file under 100MB bandwidth costs about 144 seconds. For the private network version on the intranet, with 200MB bandwidth, it only takes about 67 seconds to upload a 1GB file to our platform installed in the hospital.

The platform usability may decrease due to the relatively long reconstruction time of some optimization-based iterative reconstruction algorithms. To reduce the reconstruction time, all the algorithms run on the cloud server that is equipped with a high-performance GPU. The cloud server works in a master-slave architecture. Currently, it has one master server and two slave servers for the sake of balancing resource consumption and efficiency. The master server receives and processes service requests from the browser. When there are too many requests, the master server will forward the requests to the slave server, and call the slave server resources to process the requests collaboratively. Both master and slave servers contain an Intel Xeon Processor with 16 cores, 128GB RAM, and two NVIDIA-T4 GPUs. The whole system works in a network environment with a bandwidth of 10 Mbps. With the above hardware resources and configuration, the reconstruction time of two typical algorithms (pFISTA-SENSE [4] and PAIR [32]) on the cloud could be accelerated by about twenty-six and five times respectively, compared with the original local version (Table II).

In the image quality evaluation system, to accelerate image loading and displaying speed, a load of a batch of images is

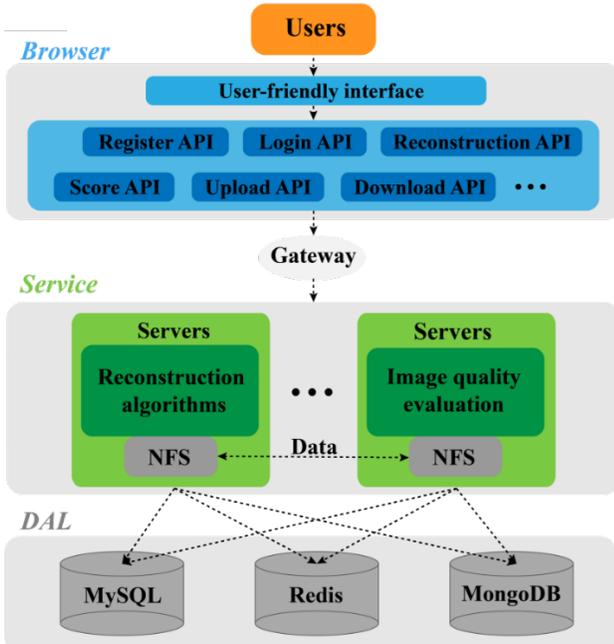

**Fig. 8.** The schematic of system architecture. Note: API, NFS, and DAL are short for application programming interface, network file system, and data access layer, respectively.

TABLE II
TRANSFER AND RECONSTRUCTION TIME OF CLOUDBRAIN-RECONAI

| Environment | Local | CloudBrain-ReconAI |
|---|---|---|
| Hardware | CPU: Intel Core i7 10700 (8 cores); RAM: 16 GB; | CPU: Intel Xeon Processor (16 cores); RAM: 128GB; GPU: NVIDIA-Tesla T4×2 |
| Upload time (s) | N. A. | 85.0 |
| Download time (s) | N. A. | 24.5 |
| Reconstruction time of pFISTA-SENSE (s) | 1984.2 | 76.3 |
| Reconstruction time of PAIR (s) | 2347.3 | 490.7 |

Note: The file for upload and download tests are 1 GB and 42 MB, respectively, and the time consumption is tested under the cloud bandwidth of 10 Mbps and local bandwidth of 100 Mbps. The matrix size of raw data for pFISTA-SENSE and PAIR are 320 × 320 × 4 × 40 (Readout × Phase encoding × Channel × Slice) and 160 × 160 × 4 × 4 × 30 (Readout × Phase encoding × Channel × Shot × Slice), respectively.

employed before displaying. This method reduces the delay of image switching effectively.

*C. Security and Privacy*

Data security and privacy of data transmission and storage are other important issues for CloudBrain-ReconAI.

For network transmission security, the platform employs encrypted transmission by JSON Web Token (JWT) authentication, Secure Sockets Layer (SSL) data encryption, and data integrity verification by MD5. JWT assigns a token containing encrypted user information and signature to the user when logging in. All subsequent requests from the user must carry this token to prove their identity. SSL verifies the server identity by a certificate authority, thus building a secure data transmission channel between the users and the server. After the data transmission, the MD5 of data will be checked to ensure the integrity of the file.

For storage security and data privacy protection, comprehensive measures have been taken on the platform. Firstly, all uploaded files will be renamed according to the MD5 code to avoid filenames that may contain user privacy. Then, the uploaded files will be doubly encrypted. The data will be encrypted by the commonly used symmetric encryption algorithm Advanced Encryption Standard (AES). Then, the key generated from AES will be encrypted further with the asymmetric encryption algorithm Rivest-Shamir-Adleman (RSA). It only takes about 11.7 seconds to encrypt and 19.7 seconds to decrypt a 3 GB file. Second, patient private information in DICOM files will be removed. Finally, all user-uploaded data will be automatically deleted after 72 hours.

IV. EXPERIMENTS

In this section, we show the usage of the CloudBrain-ReconAI through four examples of image reconstruction in fast imaging. Each starts with deep learning training, followed by image scoring and analysis. All deep learning models were trained offline and then deployed on the cloud. The reader studies were conducted with three experienced radiologists (with 13, 19, and 28 years of experience in analyzing clinical images). They were invited to score image quality through their independent accounts. All submitted scores will be automatically collected. Then, comprehensive statistical charts were generated from these scores, including a two-sample T-test [41], the Wilcoxon signed-rank test [42], the percentage plot, and the box plot.

*A. 1D Cartesian Undersampling of 2D Imaging*

Three deep learning methods, including pFISTA-NET, DOTA, and VarNET, and one compressed sensing method,

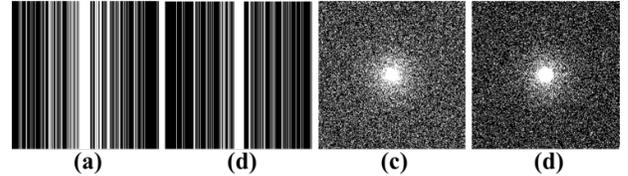

**Fig. 9.** Undersampling patterns on Cartesian grids. (a) and (b) are the 1D Cartesian pattern with 33% and 25% sampling rates, respectively. (c) and (d) are 2D random pattern with 33% and 25% sampling rates, respectively.

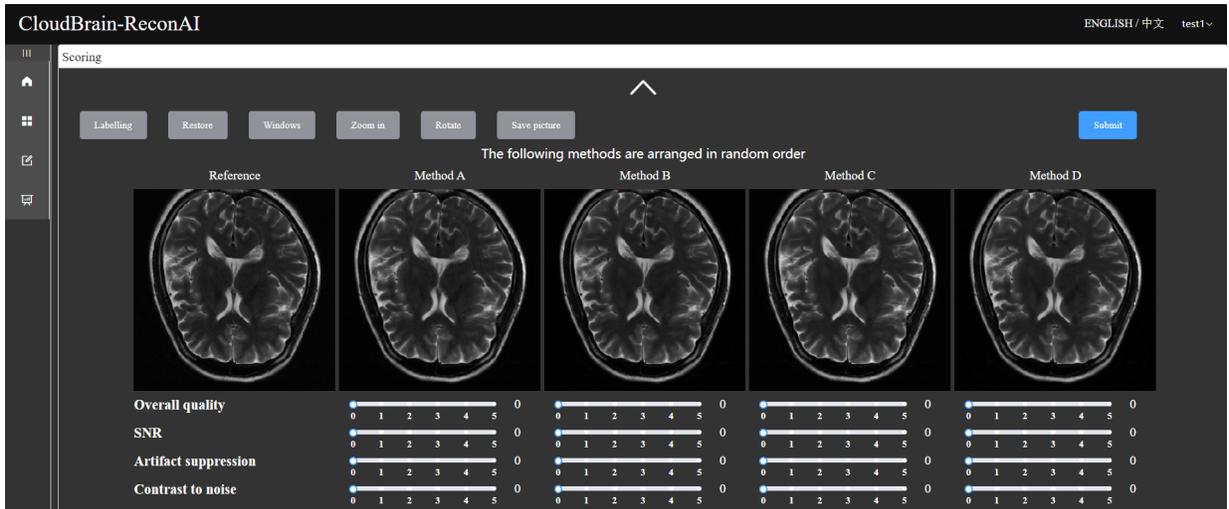

**Fig. 10.** Online reconstructed brain imaging data (ISMRM Raw Data format) on CloudBrain-ReconAI under 1D Cartesian undersampling with 33% k-space data. The methods are in random order.

pFISTA-SENSE, were compared. In the training stage, all network weights were initialized by Xavier and trained by the ADAM optimizer with a learning rate of 0.001 and exponential decay of 0.99. The training data are the fully sampled k-space data of T2-weighted brain imaging [44], which were originally fully sampled and publicly shared. This dataset has 34 subjects, each with 16 slices, and the k-space matrix size of each slice is 320×320×6 (ky×kx×coils). Fully sampled images were adopted as training labels (ideally expected output of network). Undersampled k-space data were obtained through retrospectively undersampling of the fully sampled ones according to two 1D Cartesian undersampling patterns (Figs. 9(a) and 9(b)). Reconstructed images by four methods are shown in random and blind order to radiologists (Fig. 10). After the online scoring image quality by three radiologists, statistics of scoring are automatically computed (Fig. 11).

The boxplot (Fig. 11(a)) shows that pFISTA-SENSE obtains the same median score of the overall quality but higher scores in SNR, artifact suppression, and contrast-to-noise than pFISTA and DOTA. pFISTA-SENSE achieves the comparable median values of all quality criteria as VarNet (Fig. 11(a)) but has a smaller variation than the latter does. Even though, some outliers are presented in the pFISTA-SENSE. The better performance of pFISTA-NET and VarNet is confirmed by more evidence of scores in the range of 4~5 in the percentage plot (Fig. 11(b)). When comparing the pFISTA-SENSE unrolled deep learning method, pFISTA-NET, with other deep learning methods DOTA and VarNET, both significance tests (Figs. 11(c) and (d)) indicate that pFISTA-NET has great significance with DOTA on all image quality criteria but no significance with VarNET in overall quality, SNR, and contrast-to-noise.

### B. 2D Cartesian Undersampling of 3D Imaging

The deep learning method pFISTA-NET was compared with the compressed sensing method pFISTA-SENSE. In the training stage, all network weights were initialized by Xavier and trained by the ADAM optimizer with a learning rate of 0.001 and exponential decay of 0.99. The training data are a public 3D knee dataset [45, 46] which was originally fully sampled and then retrospectively undersampled. Thus, the ground-truth images can be used as the reference in subjective evaluations. This dataset was acquired from a GE 3.0T MRI with imaging parameters that include: fast spin echo sequence, image matrix size: 320×320, coils number: 8, TR/TE: 1550/25 ms, FOV: 160×160×153.6 mm$^3$, flip angle: 90$^o$, number of slices: 256. Two 2D random sampling rates, 33% (Fig. 9 (a)) and 25% (Fig. 9 (b)), are used in the image reconstruction.

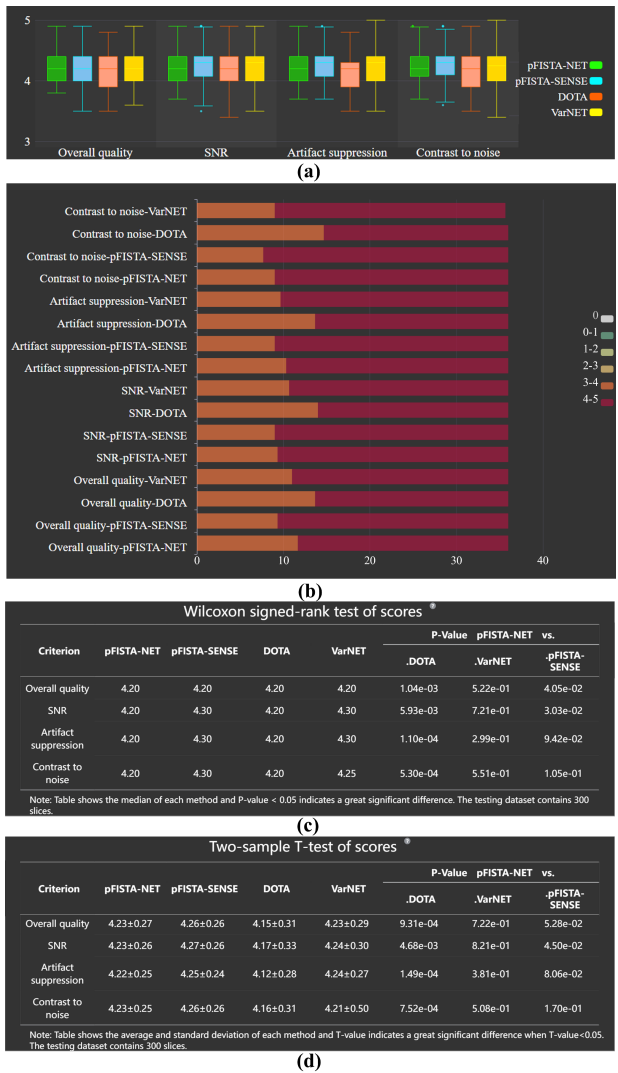

**Fig. 11.** 1D Cartesian undersampling reconstruction scoring results. (a) Box plots, (b) Percentile paragraph, (c) Wilcoxon signed-rank test of scores, (d) Two-sample T-test of scores.

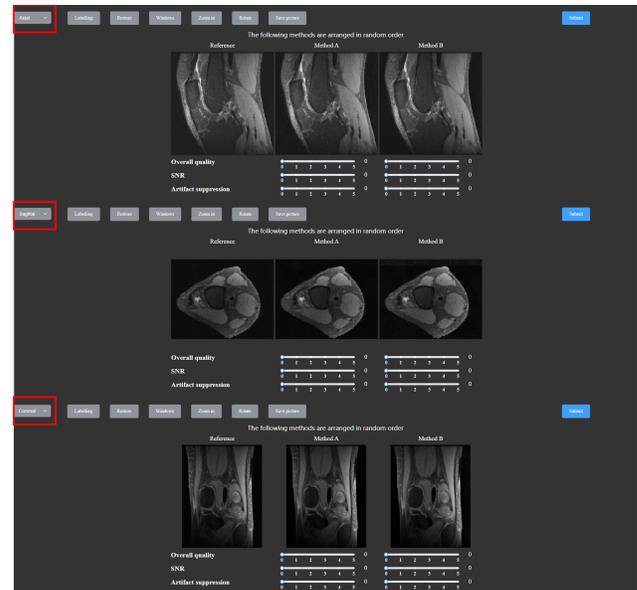

**Fig. 12.** Online reconstructed 3D imaging data (ISMRM Raw Data format) on CloudBrain-ReconAI under a 2D random pattern undersampling with 33% k-space data, and their views in axial, sagittal, and coronal plane, respectively.

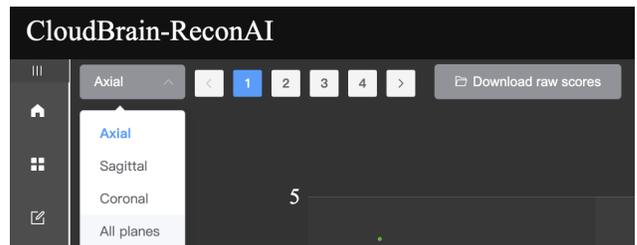

**Fig. 13.** Statistical analysis switch button for multi-views in 3D imaging.

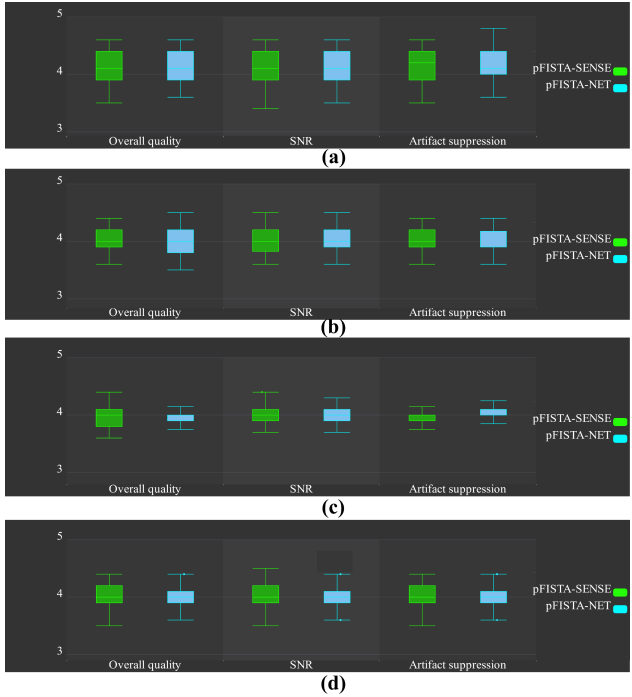

**Fig. 14.** Evaluation scores for 3D imaging. (a)-(d) are the boxplot results of axial, sagittal, coronal, and all planes, respectively.

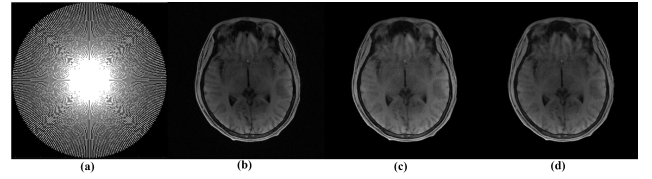

**Fig. 15.** Radial sampling and reconstructed brain imaging data (Philips raw data) under mismatched training. (a) Pseudo golden angle radial sampling trajectory of 135 spokes, (b-d) are the fully sampled image, reconstructed images by pFISTA-AI and pFISTA, respectively. Note: pFISTA-AI was trained on T2-weighted images acquired on Siemens 3.0 T scanner and then applied to reconstruct T1-weighted images acquired on a Philips 3.0T scanner.

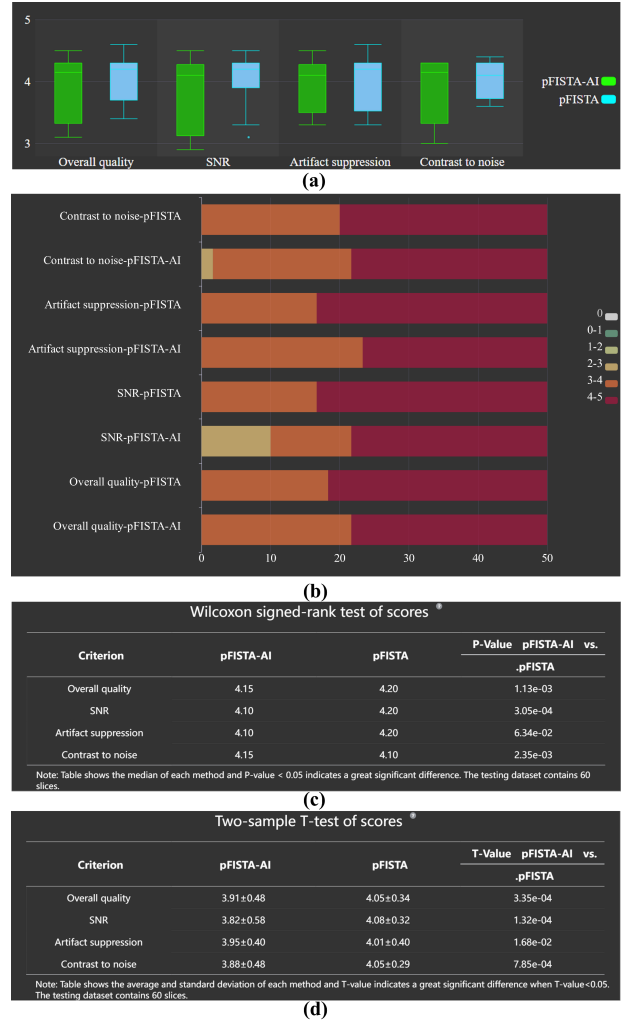

**Fig. 16.** 2D radial sampling reconstruction scoring results. (a) Box plots, (b) Percentile paragraph, (c) Wilcoxon signed-rank test of scores, (d) Two-sample T-test of scores.

Fig. 12 shows reconstructed images in three views, which can be switched by clicking the button in the upper left corner. Accordingly, statistical analysis for each view and all views have been designed (Fig. 13). The boxplot (Fig. 14) implies that pFISTA-NET is comparable with pFISTA-SENSE.

### C. 2D Radial Sampling

For 2D radial sampling, the deep learning method pFISTA-AI was compared with the compressed sensing method pFISTA. The network weights were initialized by Xavier and trained by the RADAM [47] optimizer with a learning rate of 0.0001 and exponential decay of 0.99. pFISTA-AI was trained on simulated pseudo golden angle radial sampling of the axial T2-weighted brain imaging data, which are shared at https://fastmri.med.nyu.edu and were acquired from 3.0 T MRI Siemens scanners on Cartesian grids [48]. The target reconstructed data was T1-weighted brain imaging acquired on a Philips Ingenia CX 3.0 T scanner following the pseudo golden angle radial sampling trajectory. Imaging parameters include: 3D Vane XD sequence, image matrix size: 256×256, coils number: 32, TR/TE: 3.8/1.5 ms, FOV: 256×256 mm$^2$, slice thickness: 4 mm, slice number: 41, the number of spokes in each slice: 403, radial percentage: 157%. The 403 radial spokes were adopted as the reference (the fully sampled data) and 135 spokes were used in undersampled image reconstruction. Thus, this is a mismatched reconstruction since the image contrasts of the training and target data are different.

Reconstructed images in Fig. 15 show that pFISTA-AI introduces incorrect contrast due to the mismatched training. On the contrary, the compressed sensing approach, pFISTA, does not depend on the training data, and thus produces much better images. Both the boxplot (Fig. 16(a)) and percentage plot (Fig. 16(b)) indicate that pFISTA achieves higher scores on overall quality, SNR, and artifact suppression, except for the contrast-to-noise. Besides, the Wilcoxon signed rank test (Fig. 16(c)) and Two-sample T-test (Fig. 16(d)) imply that pFISTA and pFISTA-AI have a significant difference (P < 0.05) on overall quality and SNR.

*D. Spatial-temporal Radial Sampling*

For spatial-temporal radial sampling, the deep learning method stDLNN was compared with the compressed sensing method pFISTA. The network weights were initialized by Xavier and trained by the ADAM optimizer with a learning rate of 0.001 and exponential decay of 0.99. The stTDLNN was trained on the k-space data of the lung DCE-MRI dataset [28, 29]. Both the training (28 patients) and test (2 patients) data were acquired from a Siemens MAGNETOM TimTrio 3.0T MRI [28, 29]. Each patient has 22 slices (the first 2 slices were discarded), and each slice has 66 temporal frames. Imaging parameters include: golden-angle stack-of-stars pulse sequence, image matrix size: 256×256×24, coils number: 3, TR/TE: 3.40/1.64 ms, FOV: 320×320×120 mm$^2$, slice thickness: 5 mm, slice partial Fourier: 90%, flip angle: 12º, total spokes per slice: 3000, scan time: 281 seconds. Labeled DCE images of the network (ideally expected network output) were reconstructed by the classic method iGRASP [49]. 45 spokes were used in the undersampled image reconstruction.

Three frames of the reconstruction (Fig. 17) show that lesions are clearly preserved although undersampling is performed. In addition to the lesion visibility, other subjective evaluation scores in the boxplot (Fig. 18(a)) and the percentage plot (Fig. 18(b)) also shows that the stDLNN obviously outperforms the pFISTA. This observation was firmly supported by the two significance tests (Figs. 18(c) and (d)).

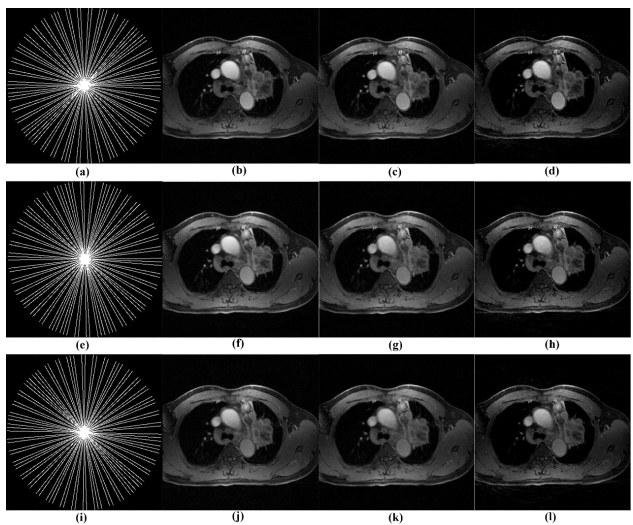

**Fig. 17.** Radial sampling and reconstructed lung DCE-MRI imaging data (Siemens raw data) under matched training. (a), (e) and (i) are golden angle radial sampling trajectories of 45 spokes of the 20$^{th}$, 40$^{th}$ and 60$^{th}$ temporal frames, respectively. (b-d), (f-h), and (j-i) are reconstructed images by iGRASP, stDLNN and pFISTA, of the 20$^{th}$, 40$^{th}$ and 60$^{th}$ temporal frames, respectively. Note: stDLNN was trained on reconstructed images using iGRASP.

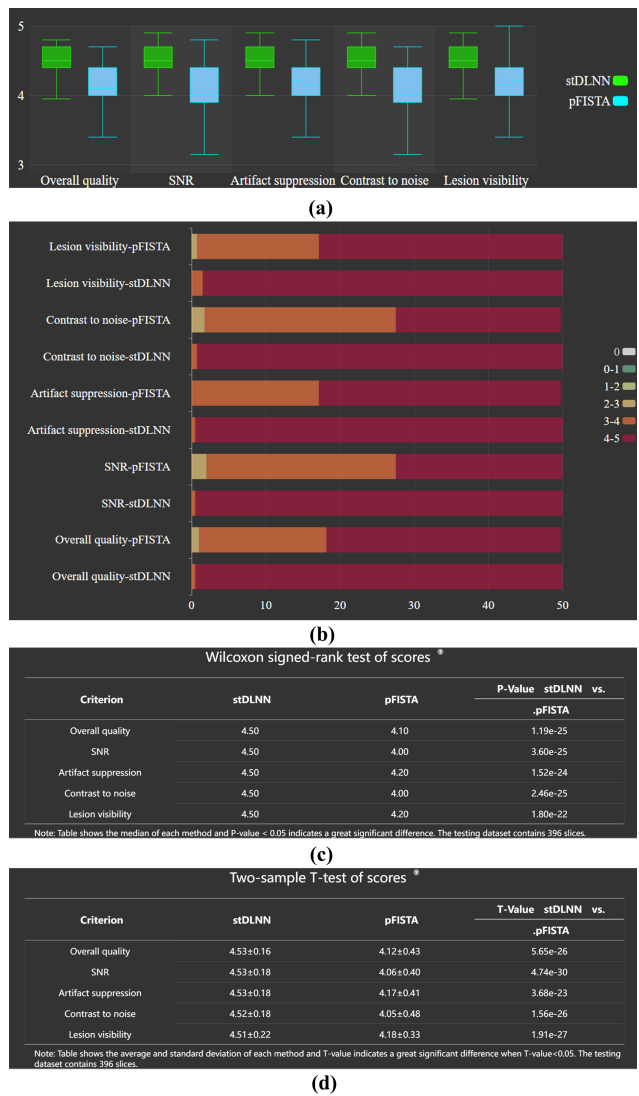

**Fig. 18.** Spatial-temporal radial sampling DCE-MRI reconstruction scoring results. (a) Box plots, (b) Percentile paragraph, (c) Wilcoxon signed-rank test of scores, and (d) Two-sample T-test of scores.

## V. DISCUSSIONS

*A. Customized Algorithm Deployment*

CloudBrain-ReconAI currently supports custom features, but it requires our platform administrators to complete this task. Algorithm developers only need to share their encrypted algorithms (compiled code) or non-encrypted algorithms (source code) and provide input/output and necessary parameters.

Customized algorithms are currently supported through six steps: 1) Developers share their algorithms via public repositories (such as GitHub) or send them to the administrator via Email quxiaobo@xmu.edu.cn. 2) Administrators ensure the algorithm runs correctly and gather the necessary algorithm parameters. 3) Administrators design the front-end interface with Vue.js [50] framework. 4) Administrators add an algorithm interface. 5) Administrators add the message queue [51] to support multi-user usage. 6) Administrators and developers verify the performance and reliability.

Future deployment of customized algorithms will follow these steps: 1) Create a standard input and output parameters for the algorithm, which will be automatically parsed and utilized during the deployment. 2) Package and share the Conda environment used by the algorithm. 3) Define "main.py" as the fixed entry point for the customized algorithm. 4) Upload to the cloud and automatically show the results.

*B. Results of DWI Reconstruction*

The PIDD in DWI reconstruction was trained on the two physic-informed synthetic data [33], which are 4-shot data with matrix sizes 160×160 and 180×180. Each dataset contains 1200 training samples generated from 120 $B_0$ DWI images (b-value=0 s/mm$^2$) with background phase and 1200 randomly generated motion phases. The $B_0$ images are acquired from 5 subjects by the 4-shot iEPI DWI sequence at a 3.0T MRI system (Philips, Ingenia CX) with 32 coils. The resolution is 1.2×1.2×5 mm$^3$. The coil number is also decreased to four by coil compression [52] to reduce computational complexity, and the sensitivity maps are estimated by ESPIRIT [53].

TABLE III
TEST DATASETS AND RECONSTRUCTION PARAMETERS

| Test dataset | 4-shot 160×160 Brain | 4-shot 180×180 Brain |
|---|---|---|
| Scanner | United Imaging, uMR890 | Philips, Ingenia CX |
| Slices | 30 | 12 |
| TR/TE (ms) | 3000/60 | / |
| B-values (s/mm$^2$) | 1000, 2000, 3000 | 1000 |
| Matrix size | 160×160 | 180×180 |
| File size (Mb) | 78.1 | 39.3 |
| Iteration (PICE) | 200 | 200 |
| Iteration (PAIR) | 100 | 100 |
| Rank (PAIR) | 30 | 30 |
| Upload time (s) | 54.62 | 28.07 |
| Recon. time (s) | 12.70/488.47/46.56 | 11.84/265.64/17.05 |

Note: Recon. time means the reconstruction time of PICE/PAIR/PIDD, respectively.

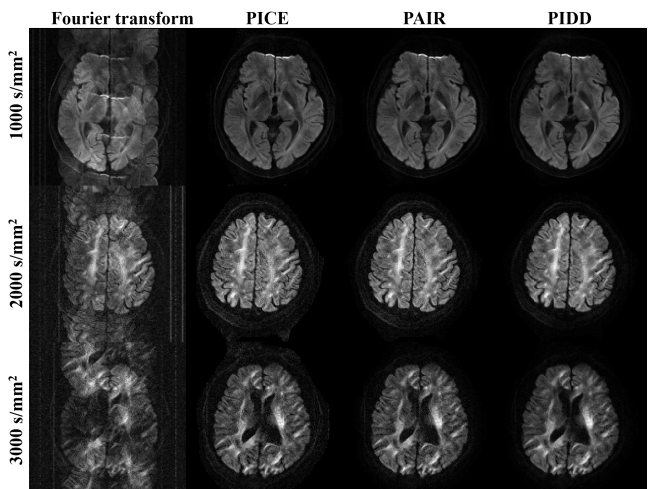

**Fig. 19.** Reconstructions of high-resolution DWI (mat file).

The target high-resolution DWI reconstruction are in vivo 4-shot DWI datasets acquired on two 3.0T MRI scanners (Table III). Before reconstruction, the N/2 Echo-Planar Imaging (EPI) ghost is corrected carefully, and sensitivity maps are estimated from the $B_0$ images.

The imaging parameters, PICE/PAIR/PIDD algorithm parameters, and the system performance, e.g. the upload and reconstruction time, are summarized in Table III. Representative reconstructions (Fig. 19) show that all algorithms can effectively remove the multi-shot image artifacts. In addition, both PAIR and PIDD have better artifact suppression ability, and PIDD provides a higher signal-to-noise on b=3000 s/mm$^2$ DWI data (3$^{rd}$ row of Fig. 19).

For the subjective evaluation, the DWI images contain 30 images with b-values 1000, 2000, and 3000 s/mm$^2$, respectively. The selected evaluation criteria for all images include overall quality, artifacts, and signal-to-noise ratio.

The boxplot (Fig. 20(a)) indicates that PIDD has the highest median scores than other methods on all criteria. The percentage plot (Fig. 20(b)) implies that the PIDD and PAIR have more scores that are distributed in the range of 4~5 than PICE. According to the Wilcoxon signed rank test (Fig. 20(c)), the results of PIDD have great significance with the other two methods (P < 0.05). In the Two-sample T-test (Fig. 20(d)), the mean of PIDD is very close to 4, indicating its great potentiality for the clinical diagnosis.

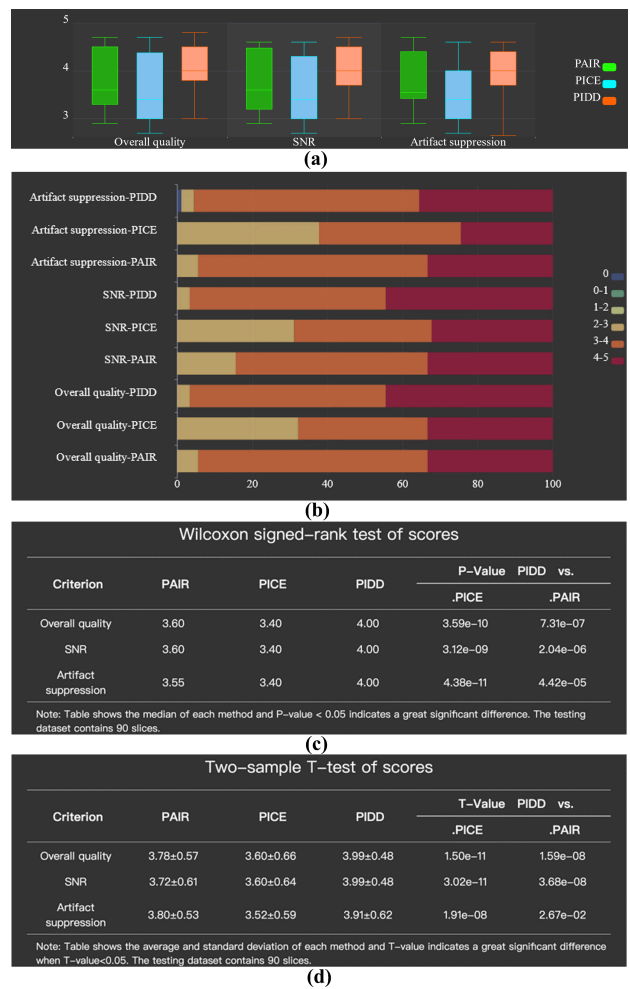

**Fig. 20.** DWI reconstruction scoring results. (a) Box plots, (b) Percentile paragraph, (c) Wilcoxon signed-rank test of scores, and (d) Two-sample T-test of scores.

## VI. CONCLUSION

We have developed the CloudBrain-ReconAI, a cloud computing and collaborative platform for online magnetic resonance image reconstruction and radiologists' subjective image evaluation. It provides easy-to-use access to the service since no software or plugin installation is required other than a common web browser. These functions are successfully demonstrated with examples of evaluating state-of-the-art approaches, including deep learning and optimization models, with applications of fast imaging and high-resolution diffusion imaging. Up to now, the CloudBrain-ReconAI has been used in multiple papers [33,54-55]. For future work, CloudBrain-ReconAI will be further improved to become a significant part of CloudBrain, an online physics-driven synthetic data [49] learning system for biomedical magnetic resonance. We expect that the CloudBrain-ReconAI will incorporate the radiologists' knowledge into developing smart reconstruction algorithms and better serve the MRI research community. The manual video can be found in the supplement file of this work or on the homepage of CloudBrain-ReconAI at https://csrc.xmu.edu.cn/CloudBrain.html.

## ACKNOWLEDGMENTS

The authors thank reviewers and editors for the valuable suggestions, which greatly improved this work. The authors also thank Yi Guo for the implementation of pFISTA-SENSE; Mengli Lu for the implementation of the pFISTA-Net; Yu Hu, Xinlin Zhang, Dichen Chen, and Haoming Fang for their helpful discussions; Mingkai Huang for the implementation of non-Cartesian algorithms; Lihua Chen for providing lung DCE-MRI dataset; Ling Qian and Ji Qi for providing cloud computing resources; and Fan Yang for revising the paper.

## REFERENCES

[1] M. Lustig, D. Donoho, and J. M. Pauly, "Compressed sensing MRI," *IEEE Signal Process. Mag.*, vol. 58, no. 6, pp. 1182-1195, 2007.

[2] M. Lustig, D. Donoho, and J. M. Pauly, "Sparse MRI: The application of compressed sensing for rapid MR imaging," *Magn. Reson. Med.*, vol. 58, no. 6, pp. 1182-1195, 2007.

[3] X. Zhang et al., "Image reconstruction with low-rankness and self-consistency of k-space data in parallel MRI," *Med. Image Anal.*, vol. 63, p. 101687, 2020.

[4] X. Zhang et al., "A guaranteed convergence analysis for the projected fast iterative soft-thresholding algorithm in parallel MRI," *Med. Image Anal.*, vol. 69, p. 101987, 2021.

[5] M. Mani, M. Jacob, D. Kelley, and V. Magnotta, "Multi-shot sensitivity-encoded diffusion data recovery using structured low-rank matrix completion (MUSSELS)," *Magn. Reson. Med.*, vol. 78, no. 2, pp. 494-507, 2017.

[6] Y. Huang et al., "Phase-constrained reconstruction of high-resolution multi-shot diffusion weighted image," *J. Magn. Reson.*, vol. 312, p. 106690, 2020.

[7] S. Wang et al., "Accelerating magnetic resonance imaging via deep learning," in *13th IEEE International Symposium on Biomedical Imaging (ISBI)*, 2016, pp. 514-517.

[8] J. Sun, H. Li, and Z. Xu, "Deep ADMM-Net for compressive sensing MRI," *Adv. Neural Inf. Process. Syst*, 2016, pp. 10-18.

[9] H. K. Aggarwal, M. P. Mani, and M. Jacob, "MoDL: Model-based deep learning architecture for inverse problems," *IEEE Trans. Med. Imaging.*, vol. 38, no. 2, pp. 394-405, 2018.

[10] H. K. Aggarwal, M. P. Mani, and M. Jacob, "MoDL-MUSSELS: Model-based deep learning for multishot sensitivity-encoded diffusion MRI," *IEEE Trans. Med. Imaging.*, vol. 39, no. 4, pp. 1268-1277, 2019.

[11] Z. Wang et al., "One-dimensional deep low-rank and sparse network for accelerated MRI," *IEEE Trans. Med. Imaging.*, vol. 42, no. 1, pp. 79-90, 2023.

[12] T. Lu et al., "pFISTA-SENSE-ResNet for parallel MRI reconstruction," *J. Magn. Reson.*, vol. 318, p. 106790, 2020.

[13] F. Wang et al., "Multiple b-value model-based residual network (MORN) for accelerated high-resolution diffusion-weighted imaging," *IEEE J. Biomed. Health Inform.*, vol. 26, no. 9, pp. 4575-4586, 2022.

[14] H. Wei, Z. Li, S. Wang, and R. Li, "Undersampled multi-contrast MRI reconstruction based on double-domain generative adversarial network," *IEEE J. Biomed. Health Inform.*, vol. 26, no. 9, pp. 4371-4377, 2022.

[15] K. Lei, A. B. Syed, X. Zhu, J. M. Pauly, and S. S. Vasanawala, "Artifact- and content-specific quality assessment for MRI with image rulers," *Med. Image Anal.*, vol. 77, p. 102344, 2022.

[16] Q. Huynh-Thu, and M. Ghanbari, "Scope of validity of PSNR in image/video quality assessment," *Electron. Lett.*, vol. 44, no. 13, pp. 800-801, 2008.

[17] A. Hore, and D. Ziou, "Image quality metrics: PSNR vs. SSIM," in *20th International Conference on Pattern Recognition*, 2010, pp. 2366-2369.

[18] M. Uecker et al., "Berkeley advanced reconstruction toolbox," in *Proc. Intl. Soc. Mag. Reson. Med*, 2015, vol. 23, no. 2486.

[19] T. Knopp and M. Grosser, "MRIReco.jl: An MRI reconstruction framework written in Julia," *Magn. Reson. Med.*, vol. 86, no. 3, pp. 1633-1646, 2021.

[20] Z. Ramzi, P. Ciuciu, and J.-L. Starck, "Benchmarking MRI reconstruction neural networks on large public datasets," *Applied Sciences.*, vol. 10, no. 5, p. 1816, 2020.

[21] M. J. Muckley et al., "Results of the 2020 fastMRI challenge for machine learning MR image reconstruction," *IEEE Trans. Med. Imaging.*, vol. 40, no. 9, pp. 2306-2317, 2021.

[22] T. Eo, H. Shin, Y. Jun, T. Kim, and D. Hwang, "Accelerating Cartesian MRI by domain-transform manifold learning in phase-encoding direction," *Med. Image Anal.*, vol. 63, no. 101689, 2020.

[23] K. Hammernik et al., "Learning a variational network for reconstruction of accelerated MRI data," *Magn. Reson. Med.*, vol. 79, no. 6, pp. 3055-3071, 2018.

[24] D. M. Hedderich et al., "Clinical evaluation of free-breathing contrast-enhanced T1w MRI of the liver using pseudo golden angle radial k-space sampling," *Rofo.*, vol. 190, no. 7, pp. 601-609, 2018.

[25] B. Qu et al., "Radial magnetic resonance image reconstruction with a deep unrolled projected fast iterative soft-thresholding network," *Comput. Biol. Med.*, vol. 168, pp. 107707, 2024.

[26] Z. Ramzi, G. Chaithya, J.-L. Starck, and P. Ciuciu, "NC-PDNet: A density-compensated unrolled network for 2D and 3D non-Cartesian MRI reconstruction," *IEEE Trans. Med. Imaging.*, vol. 41, no. 7, pp. 1625-1638, 2022.

[27] B. Qu et al., "A convergence analysis for projected fast iterative soft-thresholding algorithm under radial sampling MRI," *J. Magn. Reson.*, vol. 351, pp. 107425, 2023.

[28] L. Chen et al., "Free-breathing dynamic contrast-enhanced MRI for assessment of pulmonary lesions using golden-angle radial sparse parallel imaging," *J. Magn. Reson. Imaging.*, vol. 48, no. 2, pp. 459-468, 2018.

[29] L. Chen et al., "Improving dynamic contrast-enhanced MRI of the lung using motion-weighted sparse reconstruction: Initial experiences in patients," *Magn. Reson. Imaging.*, vol. 68, pp. 36-44, 2020.

[30] Z. Wang, H. She, Y. Zhang, and Y. P. Du, "Parallel non-Cartesian spatial-temporal dictionary learning neural networks (stDLNN) for accelerating 4D-MRI," *Med. Image Anal.*, vol. 84, pp. 102701, 2023.

[31] H. Guo et al., "POCS-enhanced inherent correction of motion-induced phase errors (POCS-ICE) for high-resolution multishot diffusion MRI," *Magn. Reson. Med.*, vol. 75, no. 1, pp. 169-180, 2016.

[32] C. Qian et al., "A paired phase and magnitude reconstruction for advanced diffusion-weighted imaging," *IEEE Trans. Biomed. Eng.*, vol. 70, no. 12, pp. 3425-3435, 2023.

[33] C. Qian et al., "Physics-informed deep diffusion MRI reconstruction: Break the bottleneck of training data in artificial intelligence," in *20th IEEE International Symposium on Biomedical Imaging* (ISBI), 2023, pp. 1-5.

[34] C. Qian et al., "Physics-informed deep diffusion MRI reconstruction: Break the bottleneck of training data in artificial intelligence," *arXiv preprint*, arXiv:2210.11388, 2024.

[35] S. J. Inati et al., "ISMRM Raw data format: A proposed standard for MRI raw datasets," *Magn. Reson. Med.*, vol. 77, no. 1, pp. 411-421, 2017.

[36] Y. Liu et al., "Projected iterative soft-thresholding algorithm for tight frames in compressed sensing magnetic resonance imaging," *IEEE Trans. Med. Imaging.*, vol. 35, no. 9, pp. 2130-2140, 2016.

[37] Q. Yang, Z. Wang, K. Guo, C. Cai, and X. Qu, "Physics-driven synthetic data learning for biomedical magnetic resonance: The imaging physics-based data synthesis paradigm for artificial intelligence," *IEEE Signal Process. Mag.*, vol. 40, no. 2, pp. 129-140, 2023.

[38] M. A. Latif et al., "Volumetric single-beat coronary computed tomography angiography: Relationship of image quality, heart rate, and body mass index. Initial patient experience with a new computed tomography scanner," *J. Comput. Assist. Tomogr.*, vol. 40, no. 5, pp. 763-772, 2016.

[39] J. Stattaus et al., "CT-guided biopsy of small liver lesions: Visibility, artifacts, and corresponding diagnostic accuracy," *Cardiovasc. Intervent. Radiol.*, vol. 30, pp. 928-935, 2007.

[40] T. Shin et al., "Diagnostic accuracy of a five-point Likert scoring system for magnetic resonance imaging (MRI) evaluated according to results of MRI/ultrasonography image-fusion targeted biopsy of the prostate," *BJU Int.*, vol. 121, no. 1, pp. 77-83, 2018.

[41] T. Suzuki et al., "Linked-color imaging improves endoscopic visibility of colorectal nongranular flat lesions," *Gastrointestinal Endoscopy.*, vol. 86, no. 4, pp. 692-697, 2017.

[42] N. Cressie, and H. Whitford, "How to use the two sample t-test," *Biom. J.*, vol. 28, no. 2, pp. 131-148, 1986.

[43] R. F. Woolson, "Wilcoxon signed-rank test," *Encyclopedia of Biostatistics*, vol. 8, 2005.

[44] K. Hammernik et al., "Learning a variational network for the reconstruction of accelerated MRI data," *Magn. Reson. Med.*, vol. 79, no. 6, pp. 3055-3071, 2018.

[45] http://mridata.org/

[46] T. Zhang, J. M. Pauly, S. S. Vasanawala, and M. Lustig, "Coil compression for accelerated imaging with Cartesian sampling," *Magn. Reson. Med.*, vol. 69, no. 2, pp. 571-582, 2013.

[47] L. Liu et al., "On the variance of the adaptive learning rate and beyond," *arXiv preprint*, arXiv:1908.03265, 2019.

[48] F. K. Zbontar et al., "FastMRI: An open dataset and benchmarks for accelerated MRI," *arXiv preprint*, arXiv: 1811.08839, 2019.

[49] L. Feng et al., "Golden-angle radial sparse parallel MRI: Combination of compressed sensing, parallel imaging, and golden-angle radial sampling for fast and flexible dynamic volumetric MRI," *Magn. Reson. Med.*, vol. 72, no. 3, pp. 707-717, 2014.

[50] C. Macrae, "Vue. js: Up and running: Building accessible and performant web apps," *O'Reilly Media, Inc,* 2018.

[51] M. S. H. Chy et al., "Comparative evaluation of Java virtual machine-based message queue services: A study on Kafka, Artemis, Pulsar, and RocketMQ," *Electronics.*, vol. 12, no. 23, p. 4792, 2023.

[52] T. Zhang, J. M. Pauly, S. S. Vasanawala, and M. Lustig, "Coil compression for accelerated imaging with Cartesian sampling," *Magn. Reson. Med.*, vol. 69, no. 2, pp. 571-582, 2013.

[53] M. Uecker et al., "ESPIRiT-an eigenvalue approach to autocalibrating parallel MRI: Where SENSE meets GRAPPA," *Magn. Reson. Med.*, vol. 71, no. 3, pp. 990-1001, 2014.

[54] Z. Wang et al., "One for multiple: Physics-informed synthetic data boosts generalizable deep learning for fast MRI reconstruction," *arXiv preprint*, arXiv:2307.13220, 2023.

[55] Z. Wang et al., "A faithful deep sensitivity estimation for accelerated magnetic resonance imaging," *IEEE J. Biomed. Health. Inform.*, vol. 28, no. 4, pp. 2126-2137, 2024.